# Prefix-Free Code Distribution Matching for 5G New Radio

Junho Cho and Ori Shental
Nokia Bell Labs
Holmdel, NJ 07733, USA
Email: {junho.cho, ori.shental}@nokia-bell-labs.com

*Abstract*—We use prefix-free code distribution matching (PCDM) for rate matching (RM) in some 5G New Radio (NR) deployment scenarios, realizing a wide range of information rates from 1.4 to 6.0 bit/symbol in fine granularity of 0.2 bit/symbol. We study the performance and implementation of the PCDM-based RM, in comparison with the low-density parity-check (LDPC)-based RM, as defined in the 5G NR standard. Simulations in the additive white Gaussian noise channel show that up to 2.16 dB gain in the signal-to-noise ratio can be obtained with the PCDM-based RM at a block error rate of $10^{-2}$ when compared to LDPC-based RM in the tested scenarios, potentially at a smaller hardware cost.

## I. INTRODUCTION

In the 5th Generation (5G) New Radio (NR) mobile broadband standard, low-density parity-check (LDPC) codes have been adopted as the channel coding scheme for user data, as recently specified in the 3rd Generation Partnership Project (3GPP) Release 15 [1]. A notable feature of the 5G NR LDPC codes is the great flexibility to support a wide range of information block lengths $K_C$, ranging from 40 to 8448 bits, and various code rates, ranging from 1/5 to 8/9 [2]–[4]. This ensures reliable transmission of user data in dynamically varying cellular channel conditions, and in various deployment scenarios where different amount of radio and hardware resources is available.

Among the many available 5G NR LDPC code parameters, finding a set of parameters to maximize the information throughput under given channel conditions and resources is a task of *rate matching* (RM). The 5G NR standard performs RM in two steps: first, coarse-grained RM chooses one of the two base graphs (BGs) and a submatrix size to lift the BG, then fine-grained RM shortens and punctures parts of the derived code in single-bit granularity. There are 51 different submatrix sizes $Z_C$ defined in the standard, in the form of $Z_C = A \times 2^j$ for $A \in \{2, 3, 5, 7, 9, 11, 13, 15\}$ and $j = 0, 1, ...$, within the range $2 \leq Z_C \leq 384$. Transmission begins with a high-rate LDPC code first, and in case the decoding fails at the receiver, incremental-redundancy hybrid automatic repeat request (HARQ) is operated such that more parity bits are transmitted for the same data until the decoding succeeds. The BGs of the 5G NR LDPC codes are made to have a special structure such that a high-rate code is always a submatrix of a lower-rate code, in order to facilitate the incremental-redundancy HARQ. Overall, the coarse- and fine-grained RM with incremental-redundancy HARQ make the number of all possible codes in an order of thousands.

Although essential to support the broad 5G NR deployment scenarios, the sheer number of LDPC codes poses a significant challenge in hardware implementation. In [5], for example, it was shown that a flexible decoder for only 12 LDPC codes (defined in the WiFi standard IEEE 802.11n/ac, with 3 different submatrix sizes and 4 code rates) consumes about 2.2× larger area than an inflexible decoder for a single code for the same throughput, when implemented on a field-programmable gate array (FPGA). In particular, multiple submatrix sizes add a greater implementation complexity than multiple code lengths, due to the intricacy associated with the design of a routing network [5]. It is therefore a daunting task to implement the whole set of 5G NR LDPC codes with as many as 51 different submatrix sizes. Moreover, this flexible coding scheme should attain up to 20 Gb/s of the downlink throughput, as required by the standard.

While RM for user data is almost solely performed by LDPC in the 5G NR standard, recent optical communication systems use *probabilistic constellation shaping* (PCS) for RM, in conjunction with a single or a few forward error correction (FEC) codes [6]. PCS shapes the probability distribution of modulation symbols such that symbols with a low energy are sent more frequently than those with a high energy, thereby reducing the average symbol energy. This implies an increased Euclidean distance between modulation symbols for the same transmit power, hence probabilistically-shaped symbols better resist the channel impairments than symbols with uniform probability distribution. Since a non-uniform distribution has a lower entropy than the uniform distribution over the same support, PCS can intrinsically adjust the information rate (IR), i.e., it can realize RM. In optical communications, PCS-based RM served as a key technology to obtain record-high spectral-efficiency transmission results in recent experiments and field trials, which led to rapid adoption in the commercial sector [6].

Motivated by the remarkable success of PCS in optical communications, we study in this work the application of PCS to mobile broadband services. We realize PCS in the probabilistic amplitude shaping (PAS) architecture [7] using *prefix-free code distribution matching* (PCDM) [8]. By transferring the role of RM to PCDM, while only a small subset of the 5G NR LDPC codes is used for FEC, we demonstrate up to 2.16 dB gain in the signal-to-noise ratio (SNR) for the same IR, at a block error rate (BLER) of $10^{-2}$. Importantly, this SNR gain may be achieved at a smaller hardware cost than the conventional LDPC-based RM, as recently shown by an FPGA implementation in optical communications scenarios [9].

TABLE I
RATE MATCHING WITH 5G LDPC CODES [1] OF LENGTH $N_C = 600$

| QAM | BG | $Z_C$ | $K_C$ | IR | QAM | BG | $Z_C$ | $K_C$ | IR | QAM | BG | $Z_C$ | $K_C$ | IR |
|---|---|---|---|---|---|---|---|---|---|---|---|---|---|---|
| 16 | 2 | 28 | 210 | 1.4 | 64 | 2 | 36 | 280 | 2.8 | 256 | 2 | 44 | 330 | 4.4 |
| 16 | 2 | 30 | 240 | 1.6 | 64 | 2 | 40 | 300 | 3.0 | 256 | 2 | 44 | 345 | 4.6 |
| 16 | 2 | 36 | 270 | 1.8 | 64 | 2 | 40 | 320 | 3.2 | 256 | 2 | 48 | 360 | 4.8 |
| 16 | 2 | 40 | 300 | 2.0 | 64 | 2 | 44 | 340 | 3.4 | 256 | 2 | 48 | 375 | 5.0 |
| 16 | 2 | 44 | 330 | 2.2 | 64 | 2 | 48 | 360 | 3.6 | 256 | 2 | 52 | 390 | 5.2 |
| 16 | 2 | 48 | 360 | 2.4 | 64 | 2 | 48 | 380 | 3.8 | 256 | 1 | 20 | 405 | 5.4 |
| 16 | 2 | 52 | 390 | 2.6 | 64 | 2 | 52 | 400 | 4.0 | 256 | 1 | 20 | 420 | 5.6 |
| 16 | 1 | 20 | 420 | 2.8 | 64 | 1 | 20 | 420 | 4.2 | 256 | 1 | 20 | 435 | 5.8 |
| 16 | 1 | 22 | 450 | 3.0 | 64 | 1 | 20 | 440 | 4.4 | 256 | 1 | 22 | 450 | 6.0 |
| 16 | 1 | 22 | 480 | 3.2 | 64 | 1 | 22 | 460 | 4.6 | | | | | |
| 16 | 1 | 24 | 510 | 3.4 | 64 | 1 | 22 | 480 | 4.8 | | | | | |
| | | | | | 64 | 1 | 24 | 500 | 5.0 | | | | | |

## II. RATE MATCHING WITH 5G NR LDPC

When a rate-$R_C$ LDPC code is used with $M^2$-ary quadrature amplitude modulation (QAM) for $M^2 \in \{4, 16, 64, 256\}$, as specified in the 5G NR standard, the achievable IR of the system is given by

$$R_{Info} = 2mR_c \tag{1}$$

in bit/symbol, where $m := \log_2 M$. This IR is said to be achieved *if* the decoding is error-free. For the 5G NR LDPC codes with incremental-redundancy HARQ, error-free decoding needs not be ensured in every transmission block, but rather a marginally low BLER (typically within the range of $10^{-3}$ to $10^{-1}$) is set as the target error performance to avoid too frequent retransmission. In this case, RM involves finding a code-modulation pair that produces the greatest $R_{Info}$ among all pairs defined in the standard such that the target BLER is fulfilled under the given channel condition. Also engaged in RM are the available radio and hardware resources in hand, and the practical requirements such as latency.

In this work, three sets of codes are selected from the 5G NR LDPC codes to produce IRs ranging from 1.4 to 6.0 bit/symbol in 0.2 bit/symbol increments to cover a wide range of channel conditions. Each set of codes has a fixed code length $N_C \in \{600, 1200, 4800\}$, which deals with a scenario with few to many resources, incurring short to long latency. For example, the extensive set of codes defined in the current 5G NR standard for the case of $N_C = 600$ is shown in Table I, where 32 different codes with 10 different submatrix sizes $Z_C$ are needed to realize the target IRs, with three different QAM orders. To support all three $N_C$ for the target IRs, 96 different LDPC codes are needed in total, with 27 different submatrix sizes.

## III. RATE MATCHING WITH PCDM

### A. PCDM

An essential component of PCS realized using the PAS architecture is the distribution matching (DM), which receives binary information bits of equal probabilities and produces modulation symbols of a target probability distribution. The transmitter of a PCS system, in the PAS architecture [7], first synthesizes a target distribution of positive real symbols using a DM, as shown in Fig. 1, then the binary representation of the positive real symbols is encoded by a binary *systematic* FEC code. The parity bits are then used as sign bits to produce real symbols that are symmetrically distributed around zero, while the systematic information bits preserve the symbol-domain probability distribution made by the DM. At the receiver side, as long as the FEC decoding recovers error-free systematic bits, the DM operation can be undone without error.

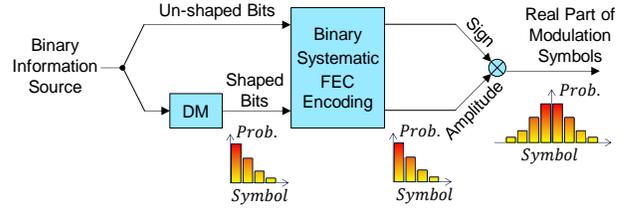

Fig. 1. PCS based on the PAS architecture [7].

TABLE II
PCDM CODE $C_2$

| Input Bits | Output Symbols |
|---|---|
| 0 | 111111 |
| 100 | 113 |
| 1010 | 111113 |
| 1011 | 11113 |
| 1100 | 1113 |
| 1101 | 1311 |
| 1110 | 3111 |
| 111100 | 133 |
| 111101 | 3113 |
| 1111100 | 1313 |
| 1111101 | 3131 |
| 1111110 | 3311 |
| 11111110 | 3133 |
| 111111110 | 3313 |
| 111111111 | 3331 |

PCDM is a method to implement DM by using *prefix-free codes* (often called *Huffman codes* [10, Ch. 5.6]). As shown in Table II, a PCDM code is constructed by concatenating two prefix-free codes, namely, binary prefix-free codewords in the left entries and non-binary (including binary) codewords in the right entries of a look-up table (LUT) in an order. A PCDM encoder reads information *bits* in a bit-by-bit manner until the first (hence shortest) matching bit sequence is found from the left entries of the LUT, then instantaneously produces a *symbol* sequence in the corresponding right entry. This variable-length bit-to-symbol encoding is repeated in an iterative manner, where each iteration starts from the first bit in the bit stream that has not been encoded yet. For example, the code in Table II (denoted by $C_2$ throughout the paper) encodes a bit stream "0 1100…" into the symbol stream "111111 1113…" Note that the right entries of $C_2$ contain only the positive real part of complex-valued 16-QAM symbols $X + iY$ for $X, Y \in \{\pm 1, \pm 3\}$, which simplifies the description and implementation. The negative real part of the symbols can be produced by using the symmetry of a probabilistic distribution around zero, as typically done in PCS systems, allowing one more information bit to be encoded as a *sign* bit in a symmetrically distributed real symbol. Generating the imaginary component is trivial; we can, for instance, use the real symbols alternately for real and imaginary components of a complex-valued QAM symbol (this approach is taken in this work). PCDM *decoding* can be described in the same manner as PCDM encoding, by changing only the role of bits and symbols, thus the details of the decoding process are omitted.

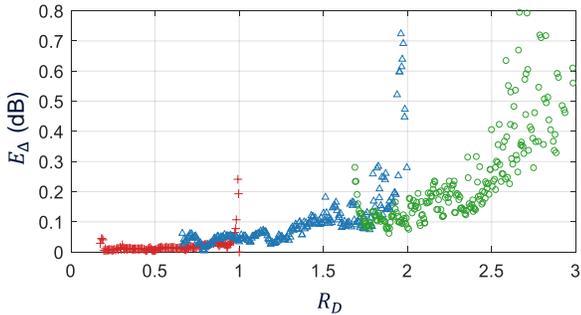

Fig. 2. Performance of PCDM codes of cardinality $|\mathcal{C}| = 24$, with real positive symbols for 16-ary (pluses), 64-ary (triangles), and 256-ary (circles) QAM.

The amount of information bits carried by each DM output symbol, called the *DM rate*, denoted by $R_D$ (in bits per positive real symbol), and the average energy $E$ of output symbols of PCDM can be easily calculated from the LUT in the limit of encoding iterations [8], assuming independent and identically distributed (IID) information bits with equal probabilities. For example, the code $\mathcal{C}_2$ realizes $R_D \approx 0.504$ with $E \approx 1.904$ asymptotically. The performance of a DM can be quantified by the *energy gap* defined as $E_\Delta \coloneqq E/E^*$, which evaluates the additional energy consumed by the DM relative to the theoretical lower limit of energy $E^*$ to achieve the same rate $R_D$. The limit $E^*$ to achieve $R_D$ is given by the average output energy of a stationary ergodic random process that generates letter $X$ from the same alphabet as the PCDM code, where $X$ follows the IID Maxwell-Boltzmann distribution [11] and produces entropy $H(X) = R_D$. The problem of constructing a good PCDM code is then to find a code that produces the smallest average energy $E$ (hence smallest $E_\Delta$) among all possible codes subject to the rate constraint $R_D \geq R_D^*$, with $R_D^*$ being the target DM rate. If we restrict the cardinality of PCDM codes (i.e., the number of rows in the LUTs) to a small number, good PCDM codes can be found by exhaustive or algorithmic search [8]. $\mathcal{C}_2$ has indeed been found in such a way, and its $E_\Delta$ is only ~0.03 dB. Note that, as per the aforementioned method in constructing a complex-valued symbol from positive real symbols, rate-$R_D$ PCDM yields $1 + R_D$ information bits per real symbol, and $2(1 + R_D)$ information bits per complex-valued symbol.

The PCDM procedure is, however, not compliant with the 5G NR standard in its current form, since it produces variable-length output at each iteration; i.e., it cannot realize fixed-rate transmission in a block-by-block manner as required by the standard. This compliance issue can be circumvented in the following manner. Namely, we use the *framing* method presented in [8], which switches the encoding method from PCDM to typical bit-to-symbol mapping for uniform QAM during the successive process. The switching position is dynamically determined from the input bit values, such that the given fixed-length bit block can be contained in a fixed-length symbol block. Framing slightly increases $E_\Delta$ in general; and the shorter the block length, the more $E_\Delta$ increases (see [8] for details). For example, the code $\mathcal{C}_2$ with $R_D \approx 0.504$ and $E \approx 1.904$ can be framed to encode an input block of length $K_D = 150$ bits in an output block of length $N_D = 300$ positive real symbols, to realize a fixed $R_D = 0.5$ in each block with a little greater average symbol energy than 1.904.

There are other known DM methods such as the constant composition DM (CCDM) [12], shell mapping (SM) [13], and enumerative sphere shaping (ESS) [14]. The CCDM needs multiplications and divisions at each iteration, making its hardware implementation very costly. The complexity of SM and ESS is much lower than CCDM, but increases with the block length. Furthermore, due to the inherently limited parallelism [14, Table 3], it is unclear if the CCDM, SM, or ESS can support 20 Gb/s of downlink throughput. There are no published papers on hardware implementation of these methods to date. On the other hand, PCDM has a low complexity, independent of the block length, and was proven through an FPGA implementation to achieve a high throughput with a massive parallelism [9], as will be discussed in Sec. III-C in more detail.

### B. Rate Matching with PCDM

We first note that the PCDM is characterized by the input and output block lengths $K_D$ and $N_D$, respectively, realizing the DM rate $R_D = K_D/N_D$ in each block, as if an LDPC code of input and output block lengths $K_C$ and $N_C$, respectively, realizes the code rate $R_C = K_C/N_C$ in each block. This already illustrates that PCDM can be used for RM, instead of the LDPC. With reference to the PAS architecture in Fig. 1, it can easily be seen that the IR of a PCS system with rate-$R_D$ DM and rate-$R_C$ coding is given by

$$R_{Info} = 2[1 + R_D - m(1 - R_C)] \quad (2)$$

(see [6], [7] for details). As a matter of fact, this shows exactly how the IR can be varied by adjusting either $R_D$ or $R_C$.

In order to perform RM with PCDM, we construct PCDM codes $\mathcal{C}$ for various $R_D$ ranging from 0.2 to 3.0 bits per positive real symbol, under the cardinality constraint $|\mathcal{C}| = 24$. There exist an enormous number of PCDM codes even with this small cardinality of 24, since the number of possible codes grows exponentially with the cardinality; e.g., for positive real symbols of 16-QAM, more than $3.4 \times 10^{11}$ different cardinality-24 codes can be constructed. Among all possible codes, the performance of the PCDM codes that have the smallest $E_\Delta$ in each $R_D$ bin of width 0.005 is shown in Fig. 2, where small $E_\Delta$ below 0.4 dB are observed across a wide range of $R_D$.

To compare PCDM- and LDPC-based RM in the considered 5G deployment scenarios, we realize the same IRs as in Sec. II using PCDM codes, in conjunction with much fewer LDPC codes than in Table I. Fixed-length framing is applied to PCDM such that each PCDM output block is mapped to exactly one LDPC code of length $N_C \in \{600, 1200, 4800\}$. This is achieved by making the PCDM output block length $N_D$ equal to $N_C/m$ for a given $N_C$. The PCDM input block length $K_D$ is then determined to meet the target IR according to (2). Shown in Table III are such determined PCDM parameters for the case of $N_C = 600$, made to be compatible with the LDPC-based RM scenario of Table I. We use 28 PCDM codes and 3 LDPC codes of 3 different submatrix sizes in Table III, one LDPC code for each QAM order (cf. top of Table III). Note, however, that it is possible to use only one submatrix size $Z_C$ for all QAM orders, if we design new LDPC codes by taking PCDM into account. More importantly, to further support the other code lengths $N_C \in \{1200, 4800\}$, we need more LDPC codes but *no* more PCDM codes, since the set of PCDM codes for $N_C = 600$ can be used for an *arbitrary* integer $N_C$ with mere change in the framing constraint, causing virtually no additional hardware cost.

TABLE III
RATE MATCHING WITH PCDM CODES AND 5G NR LDPC CODES OF LENGTH $N_C = 600$

| BG = 1 $Z_C = 20$ $N_C = 600$ $K_C = 420$ | | | | BG = 1 $Z_C = 22$ $N_C = 600$ $K_C = 480$ | | | | BG = 1 $Z_C = 24$ $N_C = 600$ $K_C = 510$ | | | |
|---|---|---|---|---|---|---|---|---|---|---|---|
| QAM | $N_D$ | $K_D$ | IR | QAM | $N_D$ | $K_D$ | IR | QAM | $N_D$ | $K_D$ | IR |
| 16 | 300 | 90  | 1.4 | 64 | 200 | 180 | 2.6 | 256 | 150 | 255 | 4.2 |
| 16 | 300 | 120 | 1.6 | 64 | 200 | 200 | 2.8 | 256 | 150 | 270 | 4.4 |
| 16 | 300 | 150 | 1.8 | 64 | 200 | 220 | 3.0 | 256 | 150 | 285 | 4.6 |
| 16 | 300 | 180 | 2.0 | 64 | 200 | 240 | 3.2 | 256 | 150 | 300 | 4.8 |
| 16 | 300 | 210 | 2.2 | 64 | 200 | 260 | 3.4 | 256 | 150 | 315 | 5.0 |
| 16 | 300 | 240 | 2.4 | 64 | 200 | 280 | 3.6 | 256 | 150 | 330 | 5.2 |
| 16 | 300 | 270 | 2.6 | 64 | 200 | 300 | 3.8 | 256 | 150 | 345 | 5.4 |
|    |     |     |     | 64 | 200 | 320 | 4.0 | 256 | 150 | 360 | 5.6 |
|    |     |     |     | 64 | 200 | 340 | 4.2 | 256 | 150 | 375 | 5.8 |
|    |     |     |     | 64 | 200 | 360 | 4.4 | 256 | 150 | 390 | 6.0 |
|    |     |     |     | 64 | 200 | 380 | 4.6 |     |     |     |     |

TABLE IV
IMPLEMENTATIONS REQUIRED FOR LDPC- AND PCDM-BASED RM WITH CODE LENGTH $N_C = 600, 1200, 4800$

| | | $N_C = 600$ | $N_C = 1200$ | $N_C = 4800$ | Total |
|---|---|---|---|---|---|
| LDPC-Based RM | # LDPC submatrix sizes | 10 | 7 | 10 | 27 |
| PCDM-Based RM | # LDPC submatrix sizes | 3 (1) | 3 (1) | 3 (1) | 9 (3) |
| | # PCDM codes | 28 | 28 | 28 | 28 |

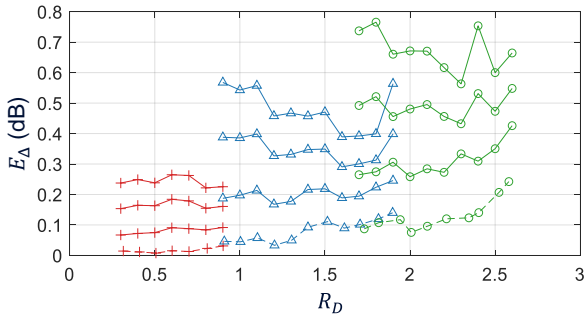

Fig. 3. Performance of PCDM codes $\mathcal{C}$ that realize $R_{Info} = 1.4, 1.6, \ldots, 6.0$ with cardinality $|\mathcal{C}| = 24$ using real positive symbols of 16-ary (pluses), 64-ary (triangles), and 256-ary (circles) QAM. The PCDM is compatible with 5G NR LDPC codes of lengths $N_C = 600$ (upper solid lines), 1200 (middle sold lines), and 4800 (lower solid lines). Also shown is the performance without a fixed-length constraint (dashed lines).

Fig. 3 shows the performance of 28 PCDM codes chosen from Fig. 2, which produce the target IRs under the framing constraints to comply with the LDPC codes of $N_C \in \{600, 1200, 4800\}$. With large PCDM block lengths $N_D$ compatible with $N_C = 4800$ (markers connected by lower solid lines), energy gap of approximately 0.1 to 0.4 dB is achieved. The energy gap increases as $N_D$ decreases, reaching almost 0.8 dB for the case of $N_C = 600$ and 256-QAM. This large gap is attributed partly to the fixed-length constraint, and partly to the cardinality constraint that becomes more prominent as the QAM order grows. However, as will be shown in Sec. IV, PCDM-based RM provides significant SNR gain even with 0.8 dB of the energy gap.

### C. Implementation Aspects of PCDM-based RM

Table IV summarizes the implementations required for LDPC- and PCDM-based RM in the 5G NR deployment scenarios with $N_C \in \{600, 1200, 4800\}$, where the numbers in the parentheses show the possible numbers if a new LDPC design criterion is applied. The PCDM-based RM uses 28 PCDM codes and 9 (3) LDPC codes of 9 (3) different submatrix sizes in total, whereas the LDPC-based RM uses 96 LDPC codes of 27 different submatrix sizes. A universal PCDM architecture is presented in [9] that can support all the 28 PCDM codes of Table IV. In this universal architecture [9], PCDM encoding is performed in a massively parallel manner, achieving 16.7 Gb/s of throughput on an FPGA platform. Moreover, to achieve the same throughput, PCDM uses substantially smaller hardware than LDPC, even with finer rate granularity [9, Sec. 4]. This shows that PCDM is a viable option to realize the fine-grained RM with the maximum throughput of 20 Gb/s, as per the 5G NR requirement.

Another important aspect is that, when PCDM performs RM, the rate of LDPC codes can be made much higher than LDPC-based RM; for example, PCDM-based RM needs $R_C \geq 0.7$ to realize all the target IRs (cf. Table III), whereas LDPC-based RM needs $R_C$ as low as 0.35 for the same IRs (cf. Table I). A higher code rate translates into a smaller number of rows in the parity-check matrix (PCM) for a fixed code length (i.e., for the same number of columns in the PCM). In case of $R_{Info} = 1.4$ bit/symbol and $N_C = 600$, the PCM for the PCDM-based RM has 44% fewer number of rows than the PCM for the LDPC-based RM, which greatly reduces the hardware cost required to implement an LDPC decoder.

## IV. PERFORMANCE EVALUATION

We evaluate the performance of the PCDM-based RM in the additive white Gaussian noise (AWGN) channel for the 5G NR deployment scenarios with $N_C = 600, 1200, 4800$, in comparison with the LDPC-based RM. For each pair of PCDM and LDPC codes, we generate $10^4$ blocks of $K_D$ IID random bits of equal probabilities, and perform PCDM encoding. Each PCDM output block is encoded into an LDPC codeword, then mapped to QAM symbols in the PAS architecture (cf. Fig. 1). After going through the AWGN channel, the received data is decoded by the belief propagation algorithm with 12 iterations. Due to the configuration of PCDM and LDPC chosen in this paper, a PCDM block error occurs if and only if an LDPC block error occurs, making the BLERs the same for the LDPC and the PCDM.

Figs. 4(a)-(c) show the IR and the SNR that is required to achieve a BLER of $10^{-2}$ with $N_C = 600, 1200, 4800$, respectively. In case of the LDPC-based RM (markers connected by dotted lines), when an IR can be achieved by multiple code-modulation pairs, a high-rate code with a low-

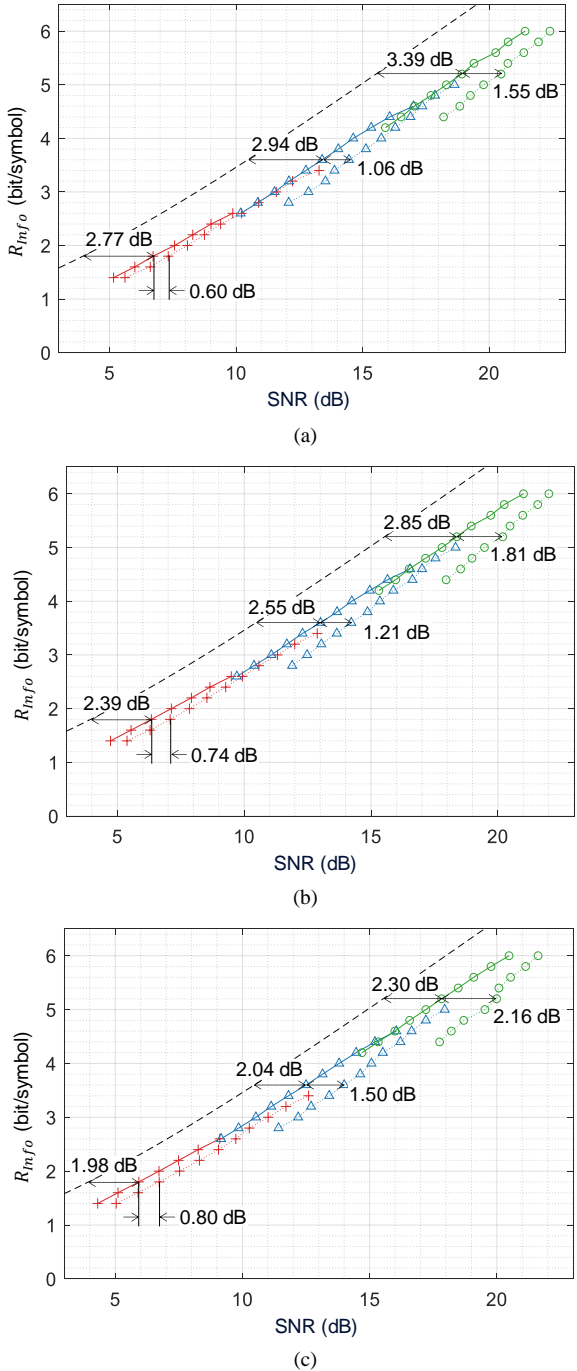

Fig. 4. IR and SNR to achieve BLER = $10^{-2}$ using LDPC- (dotted lines) and PCDM-based (solid lines) RM schemes, with (a) $N_C = 600$, (b) $N_C = 1200$, (c) $N_C = 4800$, and with 16-ary (pluses), 64-ary (triangles), and 256-ary (circles) QAM. Also shown is the AWGN channel capacity (dashed lines).

order QAM offers a greater IR than a low-rate code with a high-order QAM. The gap to the AWGN capacity (dashed lines) generally increases as the QAM order grows, as an anticipated consequence of bit-interleaved coded-modulation (BICM) with equally probable modulation symbols [15]. By contrast, when PCDM performs RM (markers connected by solid lines), the IR smoothly increases as the QAM order increases, yielding more consistent gap to the capacity than LDPC-based RM. The actual SNR gain obtained from PCDM varies with the IR and $N_C$, but significant gains are observed in a wide range of the IR, reaching up to 2.16 dB for a large $N_C$ and a large QAM order.

## V. Concluding Remarks

We studied the performance and implementation aspects of the PCDM-based RM in some 5G NR deployment scenarios. We realize a wide range of IRs from 1.4 to 6.0 bit/symbol with fine granularity of 0.2 bit/symbol, using 28 PCDM codes and only a few 5G NR LDPC codes. AWGN simulations show that up to 2.16 dB of SNR gain can be obtained with PCDM at a working point of BLER=$10^{-2}$. Furthermore, this SNR gain can potentially be achieved with a reduced hardware cost than the LDPC-based RM as currently defined in the 5G NR standard.

Although not included in the reported simulation and results, incremental-redundancy HARQ can be incorporated with PCDM. We can, for instance, use the PCDM only for the initial transmission, and transmit additional parity bits via uniform QAM if the initial transmission fails. IRs lower than 1.4 bit/symbol are not studied in this work, as it is difficult to realize them using the proposed method, but the lower IRs can be realized by using the incremental-redundancy HARQ. Full-pledged 5G NR simulations of PCDM-based RM are left for future work, which include the evaluation of the throughput with incremental-redundancy HARQ in fading channels.